\documentclass[letterpaper,10pt]{article}
\usepackage{osameet2}

\usepackage{amsmath,amssymb}
\def\CN2{\mbox{$C_N^2 \ $}}
\usepackage[pdftex,colorlinks=true,bookmarks=false,citecolor=blue,urlcolor=blue]{hyperref} %pdflatex

\begin{document}

\title{Optical Turbulence forecast: new perspectives}

\author{Elena Masciadri$^{1}$, Gianluca Martelloni$^{2}$, Alessio Turchi$^{1}$}
\address{$^{1}$INAF - Osservatorio Astrofisico di Arcetri - Largo Enrico Fermi 5, 50125, Firenze, Italy \\ $^{2}$ INSTM - Via della Lastruccia 3, Sesto Fiorentino, Firenze, Italy}
\email{e-mail: elena.masciadri@inaf.it}

\begin{abstract}
%(35-words maximum) 
In this contribution I present results achieved recently in the field of the OT forecast that push further the limit of the accuracy of the OT forecasts and open to new perspectives in this field. \end{abstract}

%\ocis{010.1330, 110.1080, 280.7060}

\section{Introduction}

The optical turbulence (OT) forecast is an extremely challenging goal with impact in many different fields. It is crucial, for example, in the context of the ground-based astronomy where the turbulence deforms the wavefront coming from the space and degrades the quality of images obtained at the focus of telescopes \cite{masciadri2013,masciadri2018}. The Adaptive Optics (AO) can, at present, correct a great part of these perturbations with very good levels of correction, particularly on high contrast imaging systems conceived to detect, for example, extrasolar planets \cite{esposito2010,fusco2014,mcintosh2014} but its efficiency strongly depends on turbulence conditions \cite{masciadri2013}. Knowing in advance the turbulence conditions is obviously important to be able to exploit the potentialities of the AO techniques. The OT forecast is also crucial in contexts different from astronomy but related to the propagation of wavefronts through the atmospheric turbulence, particularly at short wavelengths. In principle, the wavefront propagation might be not necessarily along the vertical direction (as in the astronomical case) and might be not necessarily obtained with plane wavefronts coming from sources located at infinite distances (as is the case of stars). The OT forecast is also crucial in the field of the optical communications (space-Earth) in the visible or near-infrared \cite{petit2016}. This strategy is becoming more and more attractive for the potentialities related to the faster (up to one thousand times faster) transmission and the better level of security of the optical signal with respect to the radio one \cite{robert2019}. 

The optical turbulence is characterised by different parameters such as the seeing, the isoplanatic angle, the coherence wavefront time, etc. The most relevant and common parameter is the integral value of the turbulence contained on the column of atmosphere along the line of sight called 'seeing'\footnote{In general the seeing is normalised with respect to the zenith}. 

\begin{figure}[h]
  \centering
    \includegraphics[width=5cm]{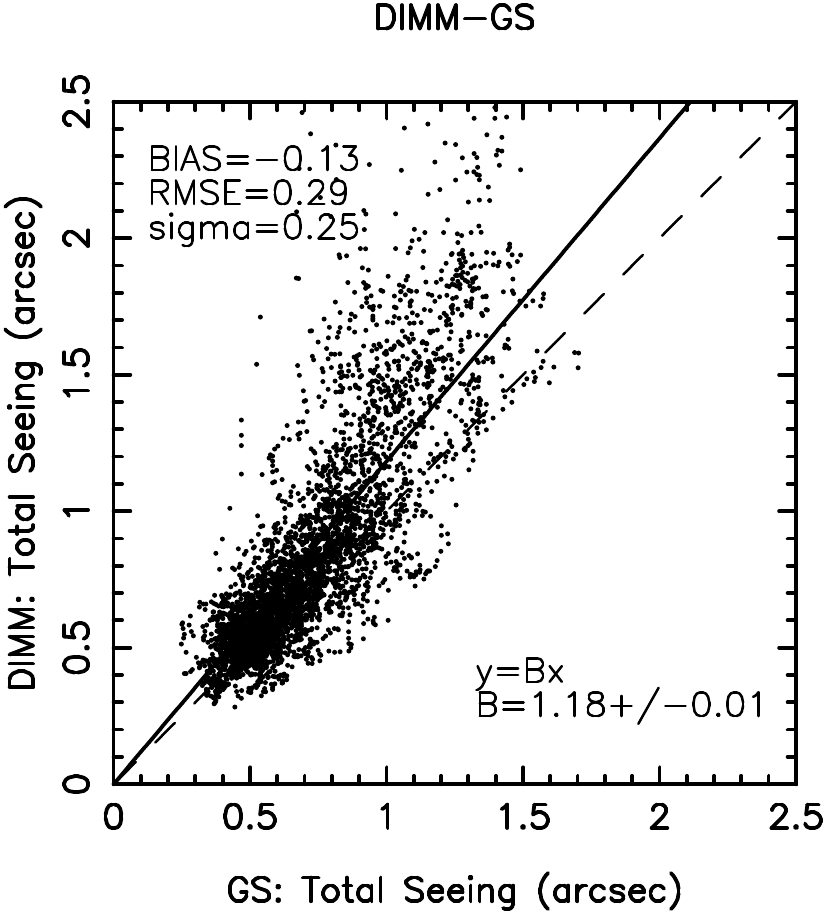}
\caption{Scattering plot of the seeing measured by two different instruments (a DIMM and a Stereo-SCIDAR) above Cerro Paranal (VLT site) on a sample of 83 nights. \label{fig0}}
\end{figure}

\section{New approaches} 

In this contribution I will deal about the new challenges and the new perspectives related to OT forecast. A not negligible difficulty for the OT forecasts is the achievable accuracy. So far evidences showed that the dispersion between forecasts of the seeing obtained with models and measured by an instrument is not so different from the dispersion obtained with two different instruments (order of [0.25 - 0.3] arcsec) \cite{masciadri2017}. This is true on a forecast time scale of the order of one day. In other words, we are considering the forecast available early in the afternoon for the next night. Fig.\ref{fig0} shows the dispersion between two instruments (DIMM \cite{sarazin1990} and Stereo-SCIDAR \cite{sheperd2014}) on a rich statistical sample of 83 nights obtained above Cerro Paranal, site of the Very Large Telescope. 

\begin{figure}[h]
  \centering
    \includegraphics[angle=90,width=14cm]{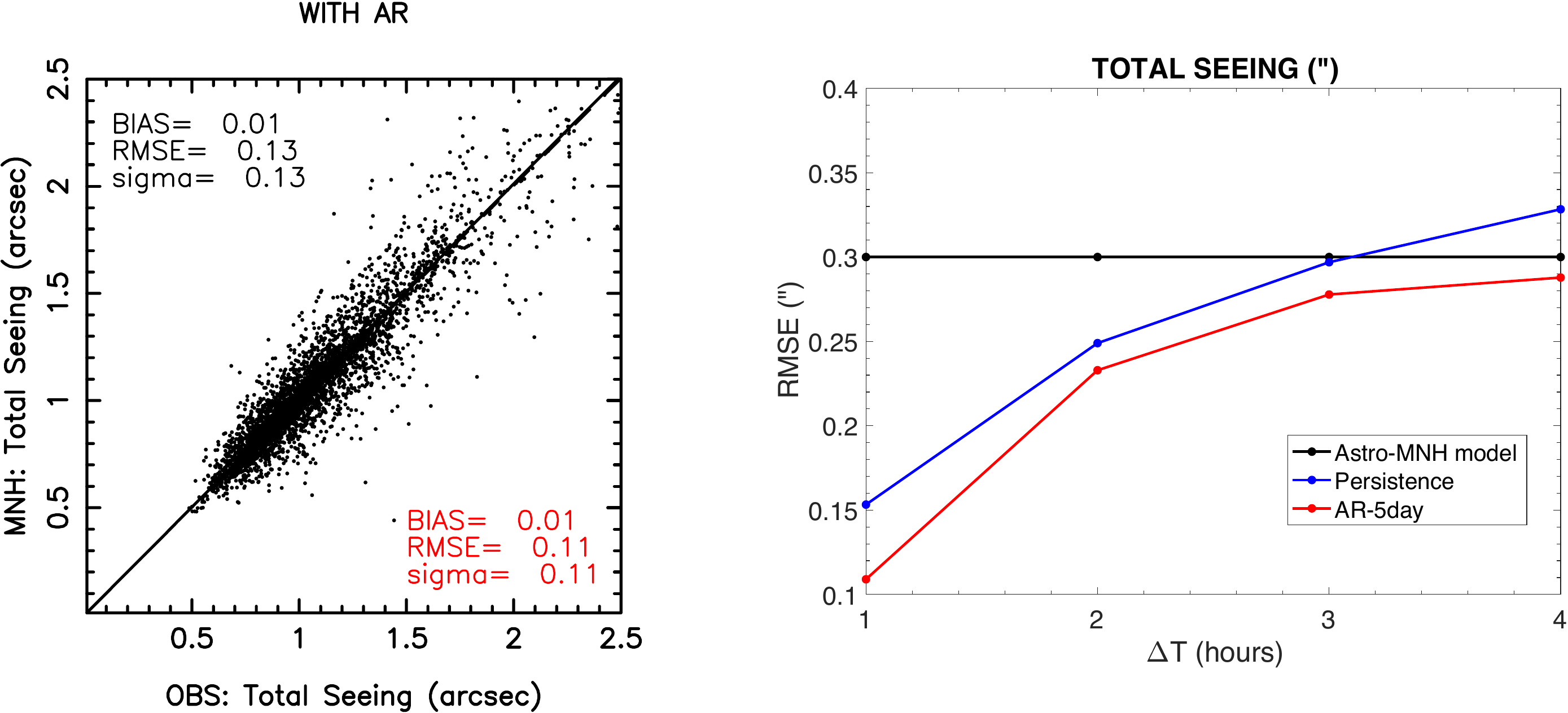}
\caption{Extracted from \cite{masciadri2020}. Left: scattering plot between observations and AR method outputs for the total seeing calculated on a sample of one year (2018). In black results considering all values, in red considering only observations below 1.5 arcsec. Right: RMSE versus the forecast time $\Delta$T (i.e the difference between the time in which the forecast is calculated and the forecast in which the forecast refers). The red line is the the RMSE obtained with the autoregressive method method applied to the forecasts done with the Astro-Meso-Nh mesoscale model. Blue line is obtained with real-time measurements and a persistence method. \label{fig1}}
\end{figure}

Recently, a new method has been proposed aiming to improve the accuracy of the forecast of the optical turbulence and the atmospherical parameters relevant for the ground-based astronomy \cite{masciadri2020}. Such an approach employs an autoregression (AR) technique. This technique takes into account, simultaneously, the forecast obtained with non-hydrostatical mesoscale atmospherical models and real-time measurements performed on a fixed and finite number of nights precedent to the present time. This method revealed to be extremely efficient in improving the forecast accuracy on short time scales (order of 1-2 hours) achieving an accuracy up to 0.1 arcsec for the seeing (see Fig.\ref{fig1}-left). A gain of a factor 3 is to be considered extremely important if we consider that in astronomical applications good and bad conditions span the [0.6 - 1.5] arcsec range. In the same study it has been put in evidence as this positive impact is even more evident from a quantitative point of view for many other atmospheric parameters relevant for the ground-based astronomy such as the temperature, wind speed and direction, the relative humidity close to the ground and the precipitable water vapour (PWV) i.e. the integral of the water vapour contained in the column of atmosphere along the zenith. This study also could show that the accuracies obtained, thanks to the method proposed, are better than what we might obtain by using only real-time measurements and a persistence method (see Fig.\ref{fig1}-right). The method \cite{masciadri2020} has been implemented in the operational forecast system called ALTA Center\footnote{\href{http://alta.arcetri.inaf.it}{http://alta.arcetri.inaf.it}} \cite{masciadri1999,masciadri2020}, a tool expressly developed to support, in operational configuration, the scheduling of the Large Binocular Telescope (LBT) observations above Mt.Graham (Arizona) \cite{veillet2016}. Such a forecasts are currently used by the science operation tool of the LBT, two telescopes of 8.4 m diameter representing a precursor of the Extremely Large Telescopes (ELTs).

\section{Conclusions and Perspectives} 

OT forecast is a crucial topic in the context of the applied research in wavefront propagation through turbulent media. Results recently achieved \cite{masciadri2020} indicate how to improve in a not negligible way the forecasts accuracy using a method that takes into account real-time measurements as well as forecasts obtained with non-hydrostatical mesoscale atmospherical models. The forecast time scale becomes a key parameter in this perspective as well as the typology of model used. Non-hydrostatical mesoscale models provide in general better performances with respect to GCM \cite{masciadri2019} on time scales for which it is reasonable to forecast the OT. Concerning the time scale, for the autoregressive method, the longer is $\Delta$T (i.e. the difference between the time in which the forecast is calculated and the time for which the forecast refers to) and the smaller is the achieved improvement on the OT forecast. After a few hours \cite{masciadri2020} the gain obviously disappears. This tells us that it is always very important to work on algorithms/methods improving OT forecast performances derived from pure numerical predictions for the long time scales.

%\footnote{\href{http://www.gemini.edu/sciops/telescopes-and-sites/locations}{http://www.gemini.edu/sciops/telescopes-and-sites/locations}}. 
%(Mauna Kea Weather Center\footnote{\href{http://mkwc.ifa.hawaii.edu}{http://mkwc.ifa.hawaii.edu}}
\section*{Acknowledgements}

We acknowledge the LBTO Director, Christian Veillet and the LBT Board for supporting us in the development of the ALTA Center Project. This work has been done in the context of the Large Binocular Telescope Observatory contract ENV001.


\begin{thebibliography}{99}

\bibitem{masciadri2013} Masciadri, E, Lascaux, F., Fini, L., "MOSE: operational forecast of the optical turbulence and atmospheric parameters at European Southern Observatory ground-based sites - I. Overview and vertical stratification of the atmospheric parameters at 0-20 km", Monthly Notices of Royal Astronomical Society, \textbf{436}, 1968--1985, (2013)

\bibitem{masciadri2018} Masciadri, E., Turchi, A., Fini, L., "Optical turbulence forecast in the Adaptive Optics realm", OSA Conference Proc. - Orlando 25-28 June 2018, https://doi.org/10.1364/3D.2018.JW5I.1

\bibitem{esposito2010} Esposito, S., Riccardi, A., Fini, L. et al., "First ligh AO (FLAO) system for LBT: final integration, acceptance test in Europe and preliminary on sky commissioning results", SPIE Proc. Adaptive Optics Systems II, \textbf{7736}, 773609-1--773609-12, (2010)

\bibitem{fusco2014} Fusco, T., Sauvage, J.F., Petit, C. et al., "Final performances and lessons learned of SAXO, the VLT-SPHERE extreme AO: from early design to on-sky results", SPIE Proc. Adaptive Optics Systems IV, \textbf{9148}, 91481U-1--91481U-15, (2014)

\bibitem{mcintosh2014} McIntosh, B., Graham, J., Ingraham, P. et al., "First light of the Gemini Planet Imager", Proc. of the National Academy of Sciences, \textbf{111}, 12661--12666, (2014)

\bibitem{petit2016} Petit, C., Vedrenne, N., Velleut, M-T, et al., "Investigation on adaptive optics performance from propagation channel characterization with the small optical transponder", Opt. Eng., \textbf{55}, 111611, (2016)

\bibitem{robert2019} Robert, C., Veilleut, M.T., Masciadri, E., et al., "Characterisation of the turbulent atmospheric channel of optical space-ground optical links with parametric models: description and cross-validation with mesoscale models and in-situ measurements", SPIE Remote Sensing and Security and Defence, Strasbourg, France, 9-12 September 2019, \textbf{11153}, 1115304-2, (2019)

\bibitem{masciadri2017} Masciadri, E., Lascaux, F., Turchi, A., Fini, L., "Optical turbulence forecast: ready for an operational application", Monthly Notices of Royal Astronomical Society, \textbf{466}, 520--539, (2017)

\bibitem{sarazin1990} Sarazin, M \& Roddier, F., "The ESO differential image motion monitor", Astron.\&Astroph., \textbf{227}, 294--300, (1990)

\bibitem{sheperd2014} Sheperd, H.W., Osborn, J., Wilson, R.W. et al., "Stereo-SCIDAR: optical turbulence profiling with high sensitivity using a modified SCIDAR instrument", Monthly Notices of Royal Astronomical Society, \textbf{437}, 3568--3577, (2014)

\bibitem{masciadri2020} Masciadri, E., Martelloni, G., Turchi, A., "Filtering technique to enhance optical turbulence performances at short time scales", Monthly Notices of Royal Astronomical Society, \textbf{492}, 140--152, (2020)

\bibitem{masciadri1999} Masciadri, E., Vernin, J., Bougeault, P., "3D mapping of optical turbulence using an atmospheric numerical model. I. A useful tool for the ground-based astronomy", Astronomy \& Astrophysics Supplement, \textbf{137}, 185--202, (1999)

\bibitem{veillet2016} Veillet, C., Ashby, D., Christou, J., Hill, J., et all., "LBTO's long march to full operation: step2", SPIE Proc. Observatory Operations: Strategies, Processes and Systems VI, \textbf{9910}, 99100S-1--99100S-14,  (2016)

\bibitem{masciadri2019} Masciadri, E., Turchi, A., Martelloni, G., "New achievements in optical turbulence forecast systems in operational mode", Conference Proc. Conference AO4ELT6 - Quebec City 9-14 June, 2019, https://arxiv.org/abs/1911.02819, (2019)

\end{thebibliography}
\end{document}